\documentclass[onecolumn]{IEEEtran}
\usepackage{amsmath}
\usepackage{xcolor}
\usepackage{amsthm}
\usepackage{float}
\usepackage{amsfonts}
\usepackage{amssymb}
\usepackage{mathrsfs}
\usepackage{graphicx}

\theoremstyle{remark}

\newcommand{\norm}[1]{\left|\left|#1\right|\right|}

\usepackage{mathrsfs}
\newfont{\Bb}{msbm10}

\newcommand{\zeroset}{\mathcal{C}_0}
\newcommand{\largeset}{\mathcal{C}_L}
\newcommand{\smallset}{\mathcal{C}_S}

\begin{document}
\title{Convergence Analysis of $l_0$-RLS Adaptive Filter}

\author{B. K. Das$^1$, {\sl Member, IEEE}, S. Mukhopadhyay$^2$, {\sl Member, IEEE}, and M. Chakraborty$^3$, {\sl Senior Member, IEEE}


\thanks{The authors are with the department of Electronics and Electrical Communication
Engineering, Indian Institute of Technology, Kharagpur, INDIA (email : $^1$ bijitbijit@gmail.com, $^2$ samratphysics@gmail.com, $^3$ mrityun@ece.iitkgp.ernet.in).} Copyright (c) 2017  BIJIT KUMAR DAS. Personal use of this material is permitted.}

\maketitle

\thispagestyle{empty}
\pagestyle{empty}

\begin{abstract}
This paper presents first and second order convergence analysis of the sparsity aware $l_0$-RLS
adaptive filter. The theorems 1 and 2 state the steady state value of mean and mean square deviation of the adaptive
filter weight vector. 
\end{abstract}
 {\bf Index terms}--Sparse Systems, Adaptive Filter, $l_0$-Recursive Least Squares,  Mean Square Deviation.

\section{Introduction}
\qquad The $l_0$-pseudonorm which measures sparsity by counting the number of nonzero elements in a vector 
can not be directly used for the regularization purpose and it is often approximated by some continuous functions, e.g., the $l_1$-norm or the absolute sum, 
the log-sum etc. Another approximation of $l_0$-norm by some exponential function has also been proposed in \cite{Mei}
and has been used to derive a new sparsity aware adaptive filter popularly called $l_0$-LMS.
This algorithm manifests excellent behavior in terms of convergence speed and steady-state mean square 
deviation for proper choice of a parameter responsible for promoting sparsity. Similarly, the conventional recursive least squares (RLS) algorithm has also
been modified to get advantage of the sparsity using $l_1$-norm penalty in \cite{Tarokh}-\cite{Tanc}, and \cite{Gian1}.
In \cite{Tanc}, $l_0$-norm regularized RLS has been proposed, and it outperforms the standard RLS and the aforementioned sparsity-aware algorithms \cite{Hero}, \cite{Mei}, \cite{Tarokh}, and \cite{Gian1}. 
In this chapter, a theoretical analysis of the $l_0$-norm regularized recursive least squares ($l_0$-RLS) is carried out.
Inspired by the work in \cite{Wang}, relevant common assumptions are taken along with some new ones and their applicability is discussed. 
Then, we also propose a combination scheme to improve the performance of the $l_0$-RLS algorithm in a scenario where signal-to-noise
(SNR) varies over time. The proposed approach combines the output of $M$ differently parameterized $l_0$-RLS adaptive filters, where the combiner
coefficients are adapted to extract the best out of the overall combination for different levels of SNR. Finally, a new sparsity promoting non-convex
function is proposed, and a sparse RLS adaptive algorithm based on this is derived.

\section{Brief Review of the $l_0$-RLS Algorithm}
\qquad In \cite{Tanc}, an RLS based sparse adaptive filter has been proposed where the cost 
function uses certain differentiable approximation of the $l_0$ norm. In particular, $||{\bf w}(n)||_0$ as
\begin{eqnarray}
  ||{\bf w}(n)||_0 \approx \sum\limits_{i=0}^{N-1} \big( 1 - exp(-\alpha|w_i(n)|)    \big),
  \label{eq:l0norm}
\end{eqnarray}
where $\alpha$ is a parameter to be chosen carefully so that if $w_i(n) \neq 0$, but $|w_i(n)|$ is small; $\alpha |w_i(n)| \approx 0$. This way, if $w_i(n)$ is zero,
or $|w_i(n)|$ is small, the corresponding factor $(1-exp(-\alpha|w_i(n)|)) \approx 0$, and if $|w_i(n)|$ is large, $(1-exp(-\alpha|w_i(n)|)) \approx 1$, which make
$ \sum\limits_{i=0}^{N-1} \big( 1 - exp(-\alpha|w_i(n)|) $ a good approximation of $||{\bf w}(n)||_0$.\\

Using the above, the RLS cost function at index $n$ is given by 
\begin{eqnarray}
\xi(n)=\sum_{m=0}^{n}\lambda^{n-m}e^2(m)+\gamma \|{\bf w}(n)\|_0,
\end{eqnarray} 
where, as before, $e(m)=d(m)-y(m),$ $m=0,1,\cdots,n$, $y(m)={\bf x}^T(m){\bf w}(m)$, $d(m) = {\bf x}^T(m){\bf w}_0 + v(m)$,
${\bf x}(m) = [x(m), x(m-1), \cdots, x(m-N+1)]^T$, $v(m)$ is an additive observation noise independent of any input $x(l) \hspace{2mm} \forall \hspace{2mm} m,l$, and
the parameter $\gamma$ is the Lagrange multiplier that controls the balance between estimation error and sparsity promoting penalty.\\
Also, the function $\|{\bf w}(n)\|_0$ is to be replaced by the abovementioned approximation.

The RLS adaptive algorithm which minimizes $\xi(n)$ and is proposed in \cite{Tanc} can be obtained as:
\begin{eqnarray}
\label{eq:evolution}
{\bf w}(n)={\bf w}(n-1)+{\bf k}(n)\epsilon(n)+\beta {\bf P}(n)g({\bf w}(n-1)),
\end{eqnarray} 
where \begin{eqnarray}
\beta=&\gamma(1-\lambda),\nonumber\\
\epsilon(n)=& d(n)-{\bf w}^T(n-1){\bf x}(n) & :\textrm{a priori error}\nonumber\\
{\bf k}(n)=&\frac{{\bf P}(n-1){\bf x}(n)}{\lambda+{{\bf x}^T(n)}{\bf P}(n-1){\bf x}(n)},\nonumber\\
{\bf P}(n)=& \frac{1}{\lambda}\left[ {\bf P}(n-1) - {\bf k}(n){\bf x}^T(n) {\bf P}(n-1)\right]\nonumber,
\end{eqnarray}
and the initialization is done as ${\bf P}(-1)=\delta{\bf I}$ ($\delta$ being a very small positive number).

The function $g({\bf w}(n-1))$ is given by $-\nabla^{s}\{ \|{\bf w}(n-1)\|_0 \}$,
where $\|{\bf w}(n-1)\|_0 $ is to be replaced by the aforementioned approximation of the $l_0$ norm.
This means $g({w}_i(n-1))$ is given by $-\frac{d}{dw_{j}(n-1)}[1-exp(-\alpha|w_{j}(n-1)|)]$, $j=0,1,\cdots, N-1$.

In \cite{Tanc}, an approximation of $g(t)$ for scalar variable $'t'$ is used which is as follows:\\
For $|t|$ large enough, $1-exp(-\alpha|t|) \approx 1$, and thus, $g(t)=0$. On the other hand, for $|t|$ small s. t. $\alpha|t| \leq 1$, 
$1 - exp(-\alpha|t|) \approx \alpha|t| - \frac{\alpha^2 t^2}{2}$, and thus $g(t) \approx \alpha^2t - \alpha sgn(t)$.

Formally, $g(t)$ is then defined as,
\begin{eqnarray}
\label{eq:define-g-function}
g(t)=\left\{\begin{array}{lr}
\alpha^2 t-\alpha\mathrm{sgn}(t), & |t|\le 1/\alpha,\\
0, & \mathrm{elsewhere}.
\end{array}\right.
\end{eqnarray}
Note that the third term in the RHS of \eqref{eq:evolution} resulting from $g({\bf w}(n-1))$ is the \emph{zero-point attraction} term and
the range $(-1/\alpha,1/\alpha)$ is called the \emph{attraction} range.
Lastly, the notations $\zeroset$, $\smallset$, and $\largeset$ are used to indicate sets of indices for zero valued tap (i.e. exactly zero), small valued tap
(i.e. $|w_{0,i}|\leq \alpha$), and large valued tap (i.e. $|w_{0,i}| > \alpha$) of ${\bf w}_0$ respectively. 

\section{Convergence Analysis}
For both first and second order convergence analysis of the $l_0$-RLS algorithm \cite{Tanc}, we adopt the following assumptions (as also made in \cite{Wang}):
\begin{enumerate}
\item{\label{assumption:data}} The data sequence $x(n)$ is a white sequence with zero mean and variance $\sigma_{x}^{2}$ and is independent of the additive noise
sequence 
$v(n)$ which is also assumed to be a zero mean sequence.
\item{\label{assumption:independence}} The incoming sequence of vectors ${\bf x}(n)$ and the filter weight vector ${\bf w}(n)$ are independent. As we
already know, this is the \textit{Independence Assumption} (\cite{hay}, \cite{sayed-book}), and is widely adopted in adaptive filtering literature for simplified analysis of the adaptive
filtering algorithms.
\item{\label{assumption:inverse-simplification}} As in the popular adaptive filtering literature \cite{hay}, \cite{sayed-book}, the forgetting factor $\lambda$ is chosen 
\emph{sufficiently} close to $1$
so that for large $n$, ${\bf P}(n)\approx (1-\lambda){\bf R}^{-1}$ where ${\bf R}$ is the autocorrelation matrix of the incoming data sequence, i.e., ${\bf R} = E[{\bf x}(n){\bf x}^T(n)]$.
\item{\label{assumption:zero-tap-weight}} At the steady state, the $k^{th}$ tap of the adaptive filter, $w_{k}(n),\ \forall k\in \zeroset$ is Gaussian distributed. 
\item{\label{assumption:tap-sign}}  It is assumed that, for large $n$, $w_{k}(n)$ are of the same sign as that of $w_{0,k}$ for $k\in \largeset\cup \smallset$.
\item{\label{assumption:attraction}} It is also assumed that, for large $n$, $w_{k}(n)$ lies outside attraction range for $k\in \largeset$ and inside attraction 
range for $k\in \smallset \cup \zeroset$, almost surely.
\end{enumerate}

\subsection{Convergence in Mean}
The first order convergence behavior of the $l_0$-RLS algorithm is described by the following theorem:\\
\textit{\textbf{Theorem $1$}}
Under the assumptions \ref{assumption:data} - \ref{assumption:attraction} above, the deviation coefficients
${\tilde w}_{k}(n):={w}_{k}(n)-w_{0,k}$, $k=0,1,\cdots,N-1$, asymptotically converge in mean to the following:
\begin{align}
\label{eq:mean-convergence-equation}
\lim\limits_{n\rightarrow \infty}E[{\tilde w}_{k}(n)]=\left\{\begin{array}{ll}
\frac{\alpha\beta}{\beta\alpha^2 - \sigma_{x}^{2}}\lim\limits_{n\to \infty}E[sgn[w_k(n)]],& k\in \zeroset,\\
\frac{\beta}{\sigma_{x}^{2} - \beta\alpha^2} g(w_{0,k}),& k\in \smallset,\\
0, & k\in \largeset.
\end{array}
\right.
\end{align}
\begin{proof}

From \eqref{eq:evolution}, we have,
\begin{eqnarray}
 & & {\tilde {\bf w}}(n) = {\bf w}(n) - {\bf w}_0 = {\tilde {\bf w}}(n-1) + \epsilon(n){\bf k}(n) + \beta {\bf P}(n)g({\bf w}(n-1)).
\end{eqnarray}

Replacing $\epsilon(n)$ by $d(n) - {\bf x}^T(n){\bf w}(n-1) = d(n) - {\bf x}^T(n)[{\bf w}_0+{\tilde {\bf w}}(n-1)],$ and $d(n)$ by ${\bf x}^T(n){\bf w}_0 + v(n)$,
and noting from \eqref{eq:inverse-define} that ${\bf I} - {\bf k}(n){\bf x}^T(n) = \lambda {\bf P}(n){\bf P}^{-1}(n-1)$, we can write
\begin{eqnarray}
\label{eq:main-evolution}
& & {\tilde {\bf w}}(n) = \lambda {\bf P}(n){\bf P}^{-1}(n-1){\tilde {\bf w}}(n-1)\nonumber\\
& & +{\bf k}(n)v(n)+\beta{\bf P}(n)g({\bf w}(n-1)).
\end{eqnarray}

Substituting ${\bf k}(n) = {\bf P}(n){\bf x}(n)$ in \eqref{eq:main-evolution} and using the assumption \ref{assumption:inverse-simplification} for large $n$,
we obtain,
\begin{eqnarray}
\label{eq:simplified-evolution}
&{\tilde {\bf w}}(n)=\lambda{\tilde {\bf w}}(n-1) + (1-\lambda){\bf R}^{-1}{\bf x}(n)v(n) \nonumber\\
&+ \beta(1-\lambda){\bf R}^{-1}g({\bf w}(n-1)).
\end{eqnarray}

Using assumption \ref{assumption:data}, the above can be simplified further as,
\begin{eqnarray}
\label{eq:simplified-evolution}
&{\tilde {\bf w}}(n)=\lambda{\tilde {\bf w}}(n-1)+\frac{(1-\lambda)}{\sigma_{x}^{2}}{\bf x}(n)v(n)\nonumber\\
&+\frac{\beta(1-\lambda)}{\sigma_{x}^{2}}g({\bf w}(n-1)).
\end{eqnarray}

Using expectation operator on both sides of the above equation, and from the orthogonality of $x(n)$ and $v(n)$ (assumption \ref{assumption:data}),
we obtain, for large $n$,
\begin{eqnarray}
 E[{\tilde {\bf w}}(n)] =  \lambda E[{\tilde {\bf w}}(n-1)] + \frac{\beta(1-\lambda)}{\sigma_{x}^{2}}E[g({\bf w}(n-1))].
 \label{eq:simplified-evolution2}
\end{eqnarray}

To evaluate $g({\bf w}(n-1))$, we analyze its $k^{th}$ component $g(w_k(n-1))$ for the three cases: $k\in \largeset$, $k\in \smallset$, and $k\in \zeroset$ separately.\\\\

\hspace{8mm}  First consider the case for $k\in\largeset$. From the definition of the function $g(\cdot)$ as given by \eqref{eq:define-g-function} and the assumption \ref{assumption:attraction}, it follows directly that,
for large $n$, $\forall  k\in \largeset$: $g(w_{k}(n-1)) = 0$ almost surely, and thus,
\begin{eqnarray}
 E[g(w_{k}(n-1))] = 0.
 \label{eq:g_large}
\end{eqnarray}

Since $0<\lambda<1$, from \eqref{eq:simplified-evolution2}, it follows that $\lim\limits_{n\to \infty}E[{\tilde w}_k(n)]=0$, for $k \in \largeset$.

Next, for evaluating $g({w}_k(n-1))$ for $k \in \smallset$, we invoke the assumptions \ref{assumption:tap-sign} and \ref{assumption:attraction}. 
From the definition in \eqref{eq:define-g-function}, and following the approach in \cite{Wang}, it is easy to see that for large $n$ and $k \in \smallset$,
the following is satisfied almost surely:
\begin{eqnarray}
 & & g(w_{k}(n-1)) = \alpha^2 w_{k}(n-1) - \alpha\mathrm{sgn}(w_{k}(n-1)) = \alpha^2 w_{k}(n-1) - \alpha\mathrm{sgn}(w_{0,k}) \nonumber\\
 & & = \alpha^2 {\tilde{w}}_{k}(n-1) + \alpha^2 w_{0,k} - \alpha\mathrm{sgn}(w_{0,k}) = \alpha^2 {\tilde{w}}_{k}(n-1) + g(w_{0,k}).
 \label{eq:g_small1}
\end{eqnarray}

Substituting in \eqref{eq:simplified-evolution2}, and simplifying, we then obtain,
\begin{eqnarray}
 \lim\limits_{n\to \infty}E[{\tilde w}_k(n)]=\frac{\beta}{\sigma_{x}^{2} - \beta\alpha^2} g(w_{0,k}), \hspace{3mm} \textrm{for} \hspace{2mm} k \in \smallset .
\end{eqnarray}

Finally, we consider the case $k \in \zeroset.$\\
Using the definition of $g(\cdot)$ in \eqref{eq:define-g-function}, and recalling the fact that, in this case, ${\tilde{w_{k}}(n)} = {w}_k(n)$, 
it is easy to see that for large $n$, the following is satisfied almost surely:
\begin{eqnarray}
 g( w_k(n-1))  = \alpha^2 w_{k}(n-1) - \alpha \mathrm{sgn}(w_{k}(n-1)), \nonumber\\
 = \alpha^2 \tilde{w}_{k}(n-1) - \alpha \mathrm{sgn}(w_{k}(n-1)), \hspace{6mm} \textrm{for} \hspace{2mm} k \in \zeroset.
 \label{eq:g_zero}
\end{eqnarray}

Substituting in \eqref{eq:simplified-evolution2} and simplifying, we can obtain
\begin{eqnarray}
 \lim\limits_{n\to \infty}E[{\tilde w}_k(n)]=\frac{\alpha\beta}{\beta\alpha^2 - \sigma_{x}^{2}}\lim\limits_{n\to \infty}E[sgn[w_k(n)]],& k\in \zeroset.
\end{eqnarray}

\end{proof}

\section{Steady State Mean Square Performance of the $l_0$-RLS}

\textit{\textbf{Theorem $2$}}
\label{theorem:steady-msd}
Under assumptions \ref{assumption:data}-\ref{assumption:attraction} above, the steady state mean square deviation of the $l_0$-RLS adaptive filter is given by
\begin{eqnarray}
\label{eq:steady-msd}
D(\infty):= &\lim\limits_{n\rightarrow \infty}E[ \norm{{\tilde {\bf w}}(n)}^2]\nonumber\\
= & D_L(\infty) + D_S(\infty) + D_0(\infty),
\end{eqnarray}
where
\begin{eqnarray}
\label{eq:dl}
 D_L(\infty) = \lim\limits_{n\to \infty}\sum\limits_{k\in \largeset} E[ {\tilde  w}^{2}_{k}(n)] = \frac{|\largeset|(1-\lambda)\sigma_{v}^{2}}{(1+\lambda)\sigma_{x}^{2}},
\end{eqnarray}
\begin{eqnarray}
\label{eq:ds}
 D_S(\infty) = \lim\limits_{n\to \infty}\sum\limits_{k\in \smallset} E[ {\tilde  w}^{2}_{k}(n)] =  \frac{|\smallset|(1-\lambda)^2\sigma_{v}^{2}}{(1-\lambda'^2)\sigma_{x}^{2}}  + \frac{\beta'G_s}{(1-\lambda'^2)},
\end{eqnarray}
where 
$\lambda' = \sqrt{\lambda^2 + \frac{2\beta\lambda(1-\lambda)\alpha^2}{\sigma_{x}^{2}} + \frac{\beta^2(1-\lambda)^2\alpha^4}{\sigma_{x}^{4}}}$,
 $\beta' = \frac{2\beta^2\lambda(1-\lambda)}{\sigma_{x}^{2}(\sigma_{x}^{2} - \beta\alpha^2)} + \frac{\beta^2(1-\lambda)^2}{\sigma_{x}^{4}} +\frac{2\beta^3(1-\lambda)^2\alpha^2}{\sigma_{x}^{4}(\sigma_{x}^{2} - \beta\alpha^2)} $,
 and $G_s = \sum\limits_{i\in\smallset}g^2(w_{0,i})$,
 
and,
\begin{eqnarray}
\label{eq:d0}
D_0(\infty) = \lim\limits_{n\to \infty}\sum\limits_{k\in \zeroset} E[ {\tilde  w}^{2}_{k}(n)] = -|\zeroset|(b_{\omega}\omega + c_{\omega}),
\end{eqnarray}
where 
$\omega = (-b_{\omega}+\sqrt{b_{\omega}^{2}-4c_{\omega}})/2, $\\
$b_{\omega} = \frac{4\alpha\beta(1-\lambda)}{\sqrt{2\pi}(1-\lambda'^2)\sigma_{x}^{2}}(\lambda + \frac{\alpha^2\beta(1-\lambda)}{\sigma_{x}^{2}})$, \\
$c_{\omega}=-\frac{(1-\lambda)^2}{1-\lambda'^2} \left(\alpha^2\frac{\beta^2}{\sigma_{x}^{4}} + \frac{\sigma_{v}^{2}}{\sigma_{x}^{2}}\right).$

\begin{proof}
We begin by investigating the evolution of the autocorrelation matrix of the filter weight deviation vector.
 From \eqref{eq:simplified-evolution}, discarding the terms involving $v(n)$ which is orthogonal to rest of the variables, we obtain \begin{eqnarray}
\label{eq:tempo-mean-square-analysis}
E[{\tilde {\bf w}}(n){{\tilde {\bf w}}^T(n)}] & ={\bf M}_1+({\bf M}_2+{\bf M}_2^T)+{\bf M}_3+{\bf M}_4,
\end{eqnarray}
where 
\begin{eqnarray}
{\bf M}_1=&\lambda^2E[{\tilde {\bf w}}(n-1){\tilde {\bf w}}^T(n-1)],\\
{\bf M}_2=&\frac{\beta\lambda(1-\lambda)}{\sigma_{x}^{2}}E[{\tilde {\bf w}}(n-1)g({\bf w}^T(n-1))],\\
{\bf M}_3=&\frac{\beta^2(1-\lambda)^2}{\sigma_{x}^{4}}E[ g({\bf w}(n-1)) g({\bf w}^T(n-1))],\\
{\bf M}_4=&\frac{(1-\lambda)^2\sigma_{v}^{2}}{\sigma_{x}^{4}} E[{\bf x}(n){\bf x}^T(n)].
\end{eqnarray}

%
%

Taking the $k^\mathrm{th}$ diagonal element of the weight deviation autocorrelation matrix,
we obtain the corresponding evolution equation:\begin{eqnarray}
\label{eq:evolve-diag-error-covariance-mat}
 & & E[ {\tilde w}_{k}^{2}(n)]=\lambda^2E[{\tilde  w}_{k}^{2}(n-1)]+\frac{2\beta\lambda(1-\lambda)}{\sigma_{x}^{2}}E[{\tilde w}_{k}(n-1)\nonumber\\
& & g(w_{k}(n-1))] + \frac{\beta^2(1-\lambda)^2}{\sigma_{x}^{4}}E[g^2(w_{k}(n-1))] \nonumber\\
& & +\frac{(1-\lambda)^2\sigma_{v}^{2}}{\sigma_{x}^{2}}.
\end{eqnarray}

To evaluate $\lim\limits_{n\to \infty} E[{\tilde w}_{k}^{2}(n)]$ recursively using \eqref{eq:evolve-diag-error-covariance-mat}, we need to evaluate the terms
$E[{\tilde{w}}_{k}(n-1)g(w_{k}(n-1))]$ and $\ E[g^2(w_{k}(n-1))]$, 
 for each $k\in \{1,2,\cdots,\ N\}$.
\label{sec:appendix-evolve-instantaneous-zero-nonzero-tap-pow} 
First consider the case of $k \in \largeset$. From assumptions \ref{assumption:attraction} and for large $n$, as seen earlier,
$g(w_{k}(n)) = 0$ ( $\forall  k \in \largeset$) almost surely.\\

This means that for large $n$, $E[{\tilde w}_k(n-1)g(w_k(n-1))] = 0$ and $E[g^2(w_k(n-1))] = 0$. 
Substituting in \eqref{eq:evolve-diag-error-covariance-mat} and noting that $0<\lambda<1$, \eqref{eq:evolve-diag-error-covariance-mat} gives rise to the following
stable, steady state solution:
\begin{eqnarray}
 \lim\limits_{n\to \infty} E[ {\tilde w}^{2}_{k}(n)] = \frac{(1-\lambda)\sigma_{v}^{2}}{(1+\lambda)\sigma_{x}^{2}}.
\end{eqnarray}

From this, $D_L(\infty)$ as given by \eqref{eq:dl} follows trivially.\\

Next we consider $k \in \smallset$. From \eqref{eq:g_small1}, one can write,
\begin{eqnarray}
  E[{\tilde{w}}_{k}(n-1)g(w_{k}(n-1))] = \alpha^2 E[{\tilde{w}}_{k}^{2}(n-1)] +  E[{\tilde{w}}_{k}(n-1)]g(w_{0,k}), \nonumber\\
\end{eqnarray}
and
\begin{eqnarray}
  E[g^2(w_{k}(n-1))] =  \alpha^4 E[{\tilde{w}}_{k}^{2}(n-1)] + g^2(w_{0,k}) + 2\alpha^2g(w_{0,k})E[{\tilde{w}}_{k}(n-1)]. \nonumber\\
\end{eqnarray}

Substituting in \eqref{eq:evolve-diag-error-covariance-mat}, we have, for $k \in \smallset$,
\begin{eqnarray}
 E[ {\tilde w}^{2}_{k}(n)] = \lambda'^2 E[{\tilde w}^{2}_{k}(n-1)]+\frac{(1-\lambda)^2\sigma_{v}^{2}}{\sigma_{x}^{2}}\nonumber\\
+(\frac{2\beta\lambda(1-\lambda)}{\sigma_{x}^{2}} + 2\frac{\beta^2(1-\lambda)^2\alpha^2}{\sigma_{x}^{4}}) g(w_{0,k})E[ {\tilde w}_{k}(n-1)]\nonumber\\
+\frac{\beta^2(1-\lambda)^2}{\sigma_{x}^{4}} g^2(w_{0,k}),
\label{eq:w2ksmall}
\end{eqnarray}
where $\lambda' = +\sqrt{\lambda^2 + \frac{2\beta\lambda(1-\lambda)\alpha^2}{\sigma_{x}^{2}} + \frac{\beta^2(1-\lambda)^2\alpha^4}{\sigma_{x}^{4}}}$.

For large $n$, from \eqref{eq:mean-convergence-equation}, $E[{\tilde w}_k(n-1)]$ can be replaced by its steady state value $\frac{\beta}{\sigma_{x}^{2} - \beta\alpha^2}g(w_0,k).$\\
The stability of \eqref{eq:w2ksmall} will then require $\lambda'$ to be less than $1$. Since $\beta = \gamma(1-\lambda)$ where $\gamma$ is very very small ($\approx 10^{-4}$)
while $\lambda \approx 1$, and $\alpha$ is typically in the range of $50-60$, for practical signals with $\sigma_{x}^{2} \approx 1$, it is easy to see that $\lambda'^2 \approx \lambda^2$
and since $0<\lambda<1$, \eqref{eq:w2ksmall} corresponds to a stable system, with 
\begin{eqnarray}
 \lim\limits_{n\to \infty} E[ {\tilde w}^{2}_{k}(n)] = \frac{(1-\lambda)^2\sigma_{v}^{2}}{(1-\lambda'^2)\sigma_{x}^{2}}  + \frac{\beta'g^2(w_{0,k})}{(1-\lambda'^2)}, \hspace{3mm} \forall k\in \smallset,
\label{eq:w2ksmall2}
\end{eqnarray}
where $\beta' = \frac{2\beta^2\lambda(1-\lambda)}{\sigma_{x}^{2}(\sigma_{x}^{2} - \beta\alpha^2)} + \frac{\beta^2(1-\lambda)^2}{\sigma_{x}^{4}} + 2\frac{\beta^3(1-\lambda)^2\alpha^2}{\sigma_{x}^{4}(\sigma_{x}^{2} - \beta\alpha^2)}.$\\\\

From \eqref{eq:w2ksmall2}, $D_S(\infty)$ as given by \eqref{eq:ds}, follows directly.\\\\

Lastly we consider the case of $k \in \zeroset$, for which we have ${\tilde w}_k(n-1) = w_k(n-1).$ Since $\beta$ is very very small and thus $\beta\alpha^2 << \sigma_{x}^{2}$,
from \eqref{eq:mean-convergence-equation}, it is safe to assume that $\lim\limits_{n \to \infty} E[{\tilde w}_k(n)] = \lim\limits_{n\to \infty} E[w_k(n)] \approx 0, \hspace{4mm} \forall k\in \zeroset.$

Also, as per assumption \ref{assumption:zero-tap-weight}, $w_k(n)$ $\forall   k \in \zeroset$, is Gaussian distributed in the steady state.
One can then apply \textit{Price's theorem} \cite{Papo} and \eqref{eq:g_zero} to write the following:
\begin{eqnarray}
    E[{\tilde{w}}_{k}(n-1)g(w_{k}(n-1))] = \alpha^2E[w_{k}^{2}(n-1)] - \sqrt{\frac{2}{\pi}}\alpha\sqrt{E[w_{k}^{2}(n-1)]},\nonumber\\
\end{eqnarray}
and 
\begin{eqnarray}
  E[g^2(w_{k}(n-1))] = \alpha^4 E[w_{k}^{2}(n-1)] + \alpha^2 - 2\sqrt{\cfrac{2}{\pi}}\alpha^3\sqrt{E[w_{k}^{2}(n-1)]}. \nonumber\\
\end{eqnarray}

Substituting in \eqref{eq:evolve-diag-error-covariance-mat}, and using the notation $\omega = \lim\limits_{n\rightarrow \infty}\sqrt{E[w_{k}^{2}(n)]}$, we then
obtain,
\begin{eqnarray}
 \omega^2 + b_{\omega} \omega + c_{\omega} = 0,
 \label{eq:omega}
\end{eqnarray}
$b_{\omega} = \frac{4\alpha\beta(1-\lambda)}{\sqrt{2\pi}(1-\lambda'^2)\sigma_{x}^{2}}(\lambda + \frac{\alpha^2\beta(1-\lambda)}{\sigma_{x}^{2}})$, \\
$c_{\omega}=-\frac{(1-\lambda)^2}{1-\lambda'^2} \left(\alpha^2\frac{\beta^2}{\sigma_{x}^{4}} + \frac{\sigma_{v}^{2}}{\sigma_{x}^{2}}\right).$\\

The roots of \eqref{eq:omega} are given by  $\omega = \cfrac{-b_{\omega} \pm \sqrt{b_{\omega}^{2} - 4c_{\omega}}}{2}$, \hspace{2mm} $\forall k \in \zeroset$.\\
Under stability assumption, $0 < \lambda' < 1$ and thus $b_{\omega} > 0$, $c_{\omega} < 0$.
Eq.\eqref{eq:omega} then has only one positive root given by $\omega = \cfrac{-b_{\omega} + \sqrt{b_{\omega}^{2} - 4c_{\omega}}}{2}$,
which provides the steady state value of $+\sqrt{E[w_{k}^{2}(n)]}$, \hspace{3mm} $\forall   k \in \zeroset$.\\

For the above choice of $\omega$, from \eqref{eq:omega}, one can also write $\lim\limits_{n\to \infty}E[{\tilde w}_{k}^{2}(n)] = \lim\limits_{n\to \infty}E[w_{k}^{2}(n)] = \omega^2 = -b_{\omega}\omega - c_{\omega}$.
From this, $D_0(\infty)$ as given by \eqref{eq:d0} follows trivially.

\end{proof}

\end{document}